\documentclass[epj]{svjour}
\usepackage{epsfig}

\advance\topmargin -8mm
\advance\textfloatsep -0.5\baselineskip

\begin{document}

\title{Evidence for the exponential distribution of income in the USA}

\author{Adrian Dr\u{a}gulescu \and Victor M. Yakovenko}

\mail{yakovenk@physics.umd.edu \\ http://www2.physics.umd.edu/\~{}yakovenk}

\institute{Department of Physics, University of Maryland, College
  Park, MD 20742-4111, USA}

\date{http://xxx.lanl.gov/abs/cond-mat/0008305, v.1 21 August 2000, v.2
   24 September 2000}

\abstract{Using tax and census data, we demonstrate that the
  distribution of individual income in the USA is exponential.  Our
  calculated Lorenz curve without fitting parameters and Gini
  coefficient 1/2 agree well with the data.  From the individual
  income distribution, we derive the distribution function of income
  for families with two earners and show that it also agrees well with
  the data.  The family data for the period 1947--1994 fit the Lorenz
  curve and Gini coefficient $3/8=0.375$ calculated for two-earners
  families.  
\PACS{{87.23.Ge}{Dynamics of social systems} \and
  {89.90.+n}{Other topics of general interest to physicists} \and
  {02.50.-r}{Probability theory, stochastic processes, and statistics}
  } }

\maketitle

%%%%%%%%%%%%%%%%%%%%%%%%%%%%%%%%%%%%%%%%%%%%%%%%%%%%%%%%%%%%%%%%%%%%%%
\vspace{-\baselineskip}
\section{Introduction}
\label{sec:introduction}

The study of income distribution has a long history.  Pareto
\cite{Pareto} proposed in 1897 that income distribution obeys a
universal power law valid for all times and countries.  Subsequent
studies have often disputed this conjecture.  In 1935, Shirras
\cite{Shirras} concluded: ``There is indeed no Pareto Law.  It is time
it should be entirely discarded in studies on distribution''.
Mandelbrot \cite{Mandelbrot} proposed a ``weak Pareto law'' applicable
only asymptotically to the high incomes.  In such a form, Pareto's
proposal is useless for describing the great majority of the
population.

Many other distributions of income were proposed: Levy, log-normal,
Champernowne, Gamma, and two other forms by Pareto himself (see a
systematic survey in the World Bank research publication
\cite{Kakwani}).  Theoretical justifications for these proposals form
two schools: socio-economic and statistical.  The former appeals to
economic, political, and demographic factors to explain the
distribution of income (e.\ g.\ \cite{Levy}), whereas the latter
invokes stochastic processes.  Gibrat \cite{Gibrat} proposed in 1931
that income is governed by a multiplicative random process, which
results in a log-normal distribution (see also \cite{Montroll}).
However, Kalecki \cite{Kalecki} pointed out that the width of this
distribution is not stationary, but increases in time.  Levy and
Solomon \cite{Solomon} proposed a cut-off at lower incomes, which
stabilizes the distribution to a power law.

In this paper, we propose that the distribution of individual income
is given by an exponential function.  This conjecture is inspired by
our previous work \cite{DY}, where we argued that the probability
distribution of money in a closed system of agents is given by the
exponential Boltzmann-Gibbs function, in analogy with the distribution
of energy in statistical physics.  In Sec.\ \ref{sec:individual}, we
compare our proposal with the census and tax data for individual
income in USA.  In Sec.\ \ref{sec:family}, we derive the distribution
function of income for families with two earners and compare it with
the census data.  The good agreement we found is discussed in Sec.\
\ref{sec:discussion}.  Speculations on the possible origins of the
exponential distribution of income are given in Sec.\
\ref{sec:origin}.

%%%%%%%%%%%%%%%%%%%%%%%%%%%%%%%%%%%%%%%%%%%%%%%%%%%%%%%%%%%%%%%%%%%%%%
\vspace{-\baselineskip}
\section{Distribution of individual income}
\label{sec:individual}

We denote income by the letter $r$ (for ``revenue'').  The probability
distribution function of income, $P(r)$, (called the probability
density in book \cite{Kakwani}) is defined so that the fraction of
individuals with income between $r$ and $r+dr$ is $P(r)\,dr$.  This
function is normalized to unity (100\%): $\int_0^\infty
P(r)\,dr=1$.  We propose that the probability distribution of
individual income is exponential:
\begin{equation}
  P_1(r)=\exp(-r/R)/R,
\label{eq:BG}  
\end{equation}
where the subscript 1 indicates individuals.  Function (\ref{eq:BG})
contains one parameter $R$, equal to the average income: \linebreak
$\int_0^\infty r\,P_1(r)\,dr=R$, and analogous to temperature in the
Boltzmann-Gibbs distribution \cite{DY}.

\begin{figure}
  \centerline{\epsfig{file=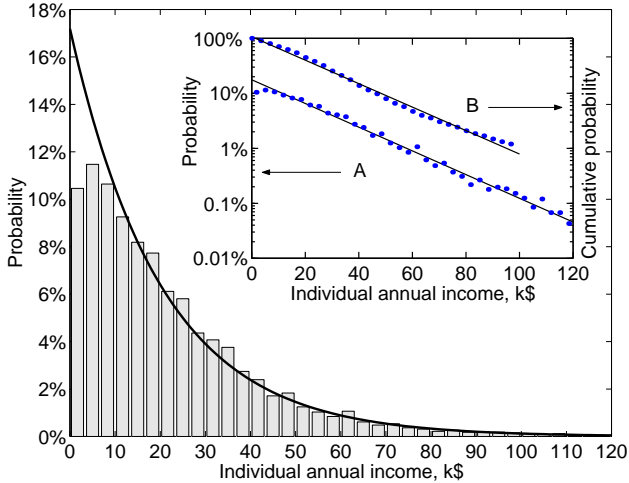,width=0.95\linewidth}}
\caption{  
  Histogram: Probability distribution of individual income from the
  U.S.\ Census data for 1996 \cite{census}.  Solid line: Fit to the
  exponential law.  Inset plot A: The same with the logarithmic
  vertical scale.  Inset plot B: Cumulative probability distribution
  of individual income from PSID for 1992 \cite{Michigan}.}
\label{fig:census}
\end{figure}

From the Survey of Income and Program Participation (SIPP)
\cite{census}, we downloaded the variable TPTOINC (total income of a
person for a month) for the first ``wave'' (a four-month period) in
1996.  Then we eliminated the entries with zero income, grouped the
remaining entries into bins of the size 10/3 k\$, counted the numbers
of entries inside each bin, and normalized to the total number of
entries.  The results are shown as the histogram in Fig.\
\ref{fig:census}, where the horizontal scale has been multiplied by 12
to convert monthly income to an annual figure.  The solid line
represents a fit to the exponential function (\ref{eq:BG}).  In the
inset, plot A shows the same data with the logarithmic vertical scale.
The data fall onto a straight line, whose slope gives the parameter
$R$ in Eq.\ (\ref{eq:BG}).  The exponential law is also often written
with the bases 2 and 10:
$P_1(r)\propto2^{-r/R_2}\propto10^{-r/R_{10}}$.  The parameters $R$,
$R_2$ and $R_{10}$ are given in line (c) of Table \ref{tab:R}.

\begin{table}[b]
  \begin{center}
  \tabcolsep 1.1ex
  \begin{tabular}{|c|lccccc|} \hline
& Source & Year & $R$ (\$) & $R_2$ (\$) & $R_{10}$ (\$) & Set size \\  \hline
a & PSID \cite{Michigan} & 1992 & 18,844 & 13,062 & 43,390 & 
  1.39$\times10^3$\\
b & IRS \cite{SailerWeber} & 1993 & 19,686 & 13,645 & 45,329 & 
  1.15$\times10^8$\\
c & SIPP$\rm_p$ \cite{census} & 1996 & 20,286 & 14,061 & 46,710 & 
  2.57$\times10^5$\\
d & SIPP$\rm_f$ \cite{census} & 1996 & 23,242 & 16,110 & 53,517 &
  1.64$\times10^5$\\
e & IRS \cite{Pub1304} & 1997 & 35,200 & 24,399 & 81,051 & 
  1.22$\times10^8$ \\  \hline
  \end{tabular}
  \end{center}
\caption{ Parameters $R$, $R_2$, and $R_{10}$ obtained by fitting data
  from different sources to the exponential law (\protect\ref{eq:BG})
  with the bases $e$, 2, and 10, and the sizes of the statistical data
  sets.}
\label{tab:R}
\end{table}

Plot B in the inset of Fig.\ \ref{fig:census} shows the data from the
Panel Study of Income Dynamics (PSID) conducted by the Institute for
Social Research of the University of Michigan \cite{Michigan}.  We
downloaded the variable V30821 ``Total 1992 labor income'' for
individuals from the Final Release 1993 and processed the data in a
similar manner.  Shown is the cumulative probability distribution of
income $N(r)$ (called the probability distribution in book
\cite{Kakwani}).  It is defined as $N(r)=\int_r^\infty P(r')\, dr'$
and gives the fraction of individuals with income greater than $r$.
For the exponential distribution (\ref{eq:BG}), the cumulative
distribution is also exponential: $N_1(r)=\int_r^\infty P_1(r')\,
dr'=\exp(-r/R)$.  Thus, $R_2$ is the median income; 10\% of population
have income greater than $R_{10}$ and only 1\% greater than $2R_{10}$.
The points in the inset fall onto a straight line in the logarithmic
scale.  The slope is given in line (a) of Table \ref{tab:R}.

\begin{figure}
\centerline{\epsfig{file=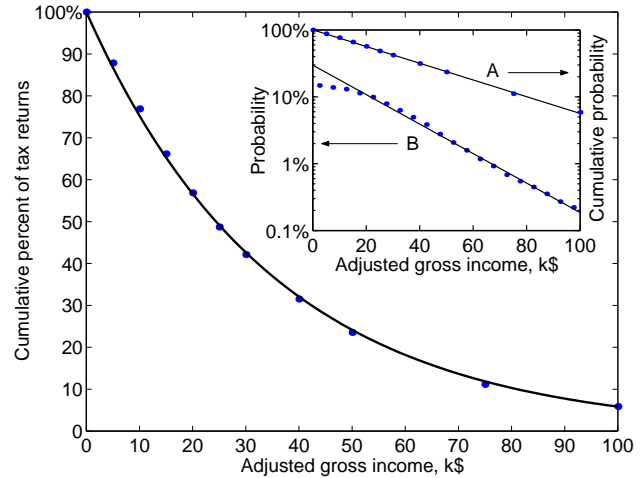,width=0.965\linewidth}}
\caption{
  Points: Cumulative fraction of tax returns vs income from the IRS
  data for 1997 \cite{Pub1304}.  Solid line: Fit to the exponential
  law.  Inset plot A: The same with the logarithmic vertical scale.
  Inset plot B: Probability distribution of individual income from the
  IRS data for 1993 \cite{SailerWeber}.}
\label{fig:IRS}
\end{figure}

The points in Fig.\ \ref{fig:IRS} show the cumulative distribution of
tax returns vs income in 1997 from column 1 of Table 1.1 of Ref.\ 
\cite{Pub1304}.  (We merged 1 k\$ bins into 5 k\$ bins in the interval
1--20 k\$.)  The solid line is a fit to the exponential law.  Plot A
in the inset of Fig.\ \ref{fig:IRS} shows the same data with the
logarithmic vertical scale.  The slope is given in line (e) of Table
\ref{tab:R}.  Plot B in the inset of Fig.\ \ref{fig:IRS} shows the
distribution of individual income from tax returns in 1993
\cite{SailerWeber}.  The logarithmic slope is given in line (b) of
Table \ref{tab:R}.

While Figs.\ \ref{fig:census} and \ref{fig:IRS} clearly demostrate the
fit of income distribution to the exponential form, they have the
following drawback.  Their horizontal axes extend to $+\infty$, so the
high-income data are left outside of the plots.  The standard way to
represent the full range of data is the so-called Lorenz curve (for an
introduction to the Lorenz curve and Gini coefficient, see book
\cite{Kakwani}).  The horizontal axis of the Lorenz curve, $x(r)$,
represents the cumulative fraction of population with income below
$r$, and the vertical axis $y(r)$ represents the fraction of income
this population accounts for:
\begin{equation}
  x(r)=\int_0^r P(r')\,dr',\quad
  y(r)=\frac{\int_0^r r' P(r')\,dr'}{\int_0^\infty r' P(r')\,dr'}.
\label{eq:xy}
\end{equation}
As $r$ changes from 0 to $\infty$, $x$ and $y$ change from 0 to 1, and
Eq.\ (\ref{eq:xy}) parametrically defines a curve in the
$(x,y)$-space.

Substituting Eq.\ (\ref{eq:BG}) into Eq.\ (\ref{eq:xy}), we find
\begin{equation}
  x(\tilde{r})=1-\exp(-\tilde{r}), \quad
  y(\tilde{r})=x(\tilde{r})-\tilde{r}\exp(-\tilde{r}),
\label{eq:xy1}
\end{equation}
where $\tilde{r}=r/R$.  Excluding $\tilde{r}$, we find the explicit
form of the Lorenz curve for the exponential distribution:
\begin{equation}
  y=x+(1-x)\ln(1-x).
\label{eq:Lorenz}
\end{equation}
$R$ drops out, so Eq.\ (\ref{eq:Lorenz}) has no fitting parameters.

\begin{figure}
\centerline{\epsfig{file=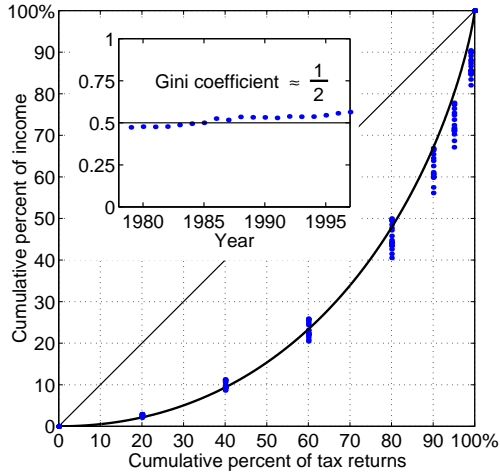,width=0.75\linewidth}}
\caption{
  Solid curve: Lorenz plot for the exponential distribution.  Points:
  IRS data for 1979--1997 \cite{Petska}.  Inset points: Gini
  coefficient data from IRS \cite{Petska}.  Inset line: The calculated
  value 1/2 of the Gini coefficient for the exponential distribution.}
\label{fig:Lorenz_1}
\end{figure}

The function (\ref{eq:Lorenz}) is shown as the solid curve in Fig.\ 
\ref{fig:Lorenz_1}.  The straight diagonal line represents the Lorenz
curve in the case where all population has equal income.  Inequality
of income distribution is measured by the Gini coefficient $G$, the
ratio of the area between the diagonal and the Lorenz curve to the
area of the triangle beneath the diagonal: $G=2\int_0^1(x-y)\,dx$.
The Gini coefficient is confined between 0 (no inequality) and 1
(extreme inequality).  By substituting Eq.\ (\ref{eq:Lorenz}) into the
integral, we find the Gini coefficient for the exponential
distribution: $G_1=1/2$.

The points in Fig.\ \ref{fig:Lorenz_1} represent the tax data during
1979--1997 from Ref.\ \cite{Petska}.  With the progress of time, the
Lorenz points shifted downward and the Gini coefficient increased from
0.47 to 0.56, which indicates increasing inequality during this
period.  However, overall the Gini coefficient is close to the value
0.5 calculated for the exponential distribution, as shown in the inset
of Fig.\ \ref{fig:Lorenz_1}.

%%%%%%%%%%%%%%%%%%%%%%%%%%%%%%%%%%%%%%%%%%%%%%%%%%%%%%%%%%%%%%%%%%%%%%
\vspace{-\baselineskip}
\section{Income distribution for two-earners families}
\label{sec:family}

\begin{figure}
  \centerline{\epsfig{file=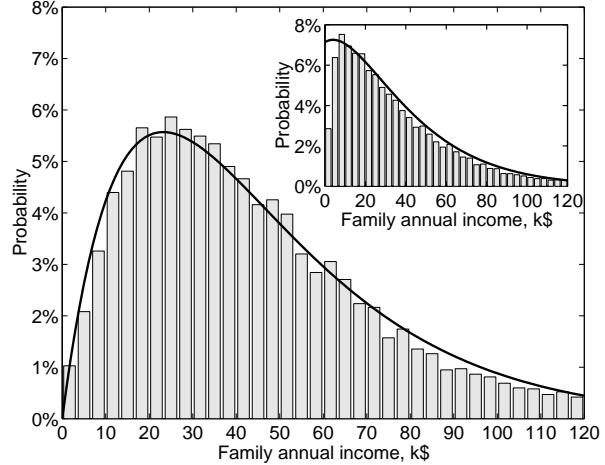,width=0.9\linewidth}}
\caption{
  Histogram: Probability distribution of income for families with two
  adults in 1996 \cite{census}.  Solid line: Fit to Eq.\ 
  (\ref{eq:family}).  Inset histogram: Probability distribution of
  income for all families in 1996 \cite{census}. Inset solid line:
  $0.45P_1(r)+0.55P_2(r)$.}
\label{fig:census_2}
\end{figure}

Now let us discuss the distribution of income for families with two
earners.  The family income $r$ is the sum of two individual incomes:
$r=r_1+r_2$.  Thus, the probability distribution of the family income
is given by the convolution of the individual probability
distributions \cite{Feller}.  If the latter are given by the
exponential function (\ref{eq:BG}), the two-earners probability
distribution function $P_2(r)$ is
\begin{equation}
  P_2(r)=\int_0^{r}P_1(r')P_1(r-r')\,dr'= \frac{r}{R^2}e^{-r/R}.
\label{eq:family}
\end{equation}
The function $P_2(r)$ (\ref{eq:family}) differs from the function
$P_1(r)$ (\ref{eq:BG}) by the prefactor $r/R$, which reflects the
phase space available to compose a given total income out of two
individual ones.  It is shown as the solid curve in Fig.\ 
\ref{fig:census_2}.  Unlike $P_1(r)$, which has a maximum at zero
income, $P_2(r)$ has a maximum at $r=R$ and looks qualitatively
similar to the family income distribution curves in literature
\cite{Levy}.

From the same 1996 SIPP that we used in Sec.\ \ref{sec:individual}
\cite{census}, we downloaded the variable TFTOTINC (the total family
income for a month), which we then multiplied by 12 to get annual
income.  Using the number of family members (the variable EFNP) and
the number of children under 18 (the variable RFNKIDS), we selected
the families with two adults.  Their distribution of family income is
shown by the histogram in Fig.\ \ref{fig:census_2}.  The fit to the
function (\ref{eq:family}), shown by the solid line, gives the
parameter $R$ listed in line (d) of Table \ref{tab:R}.  The families
with two adults and more than two adults constitute 44\% and 11\% of
all families in the studied set of data.  The remaining 45\% are the
families with one adult.  Assuming that these two classes of families
have two and one earners, we expect the income distribution for all
families to be given by the superposition of Eqs.\ (\ref{eq:BG}) and
(\ref{eq:family}): $0.45P_1(r)+0.55P_2(r)$.  It is shown by the solid
line in the inset of Fig.\ \ref{fig:census_2} (with $R$ from line (d)
of Table \ref{tab:R}) with the all families data histogram.

By substituting Eq.\ (\ref{eq:family}) into Eq.\ (\ref{eq:xy}), we
calculate the Lorenz curve for two-earners families:
\begin{equation}
  x(\tilde{r})= 1 - (1+\tilde{r}) e^{-\tilde{r}}, \;\;
  y(\tilde{r})= x(\tilde{r}) -\tilde{r}^2 e^{-\tilde{r}}/2.
\label{eq:xy2}
\end{equation}
It is shown by the solid curve in Fig.\ \ref{fig:Lorenz_2}.  Given
that $x-y=\tilde{r}^2\exp(-\tilde{r})/2$ and
$dx=\tilde{r}\exp(-\tilde{r})\,d\tilde{r}$, the Gini coefficient for
two-earners families is: $G_2=2\int_0^1(x-y)\,dx=
\int_0^\infty\tilde{r}^3\exp(-2\tilde{r})\,d\tilde{r}=3/8=0.375$.  The
points in Fig.\ \ref{fig:Lorenz_2} show the Lorenz data and Gini
coefficient for family income during 1947--1994 from Table 1 of Ref.\
\cite{history}.  The Gini coefficient is very close to the calculated
value 0.375.

%%%%%%%%%%%%%%%%%%%%%%%%%%%%%%%%%%%%%%%%%%%%%%%%%%%%%%%%%%%%%%%%%%%%%%
\vspace{-\baselineskip}
\section{Discussion}
\label{sec:discussion}

\begin{figure}
  \centerline{\epsfig{file=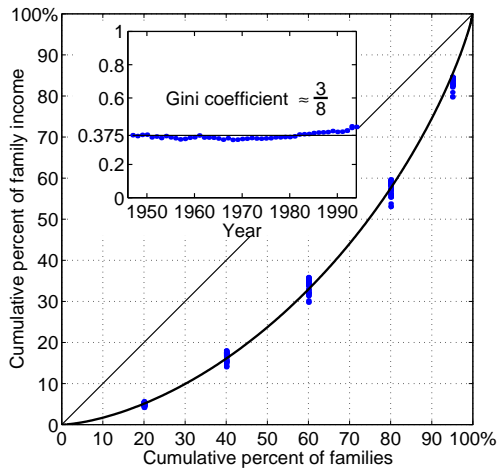,width=0.75\linewidth}}
\caption{
  Solid curve: Lorenz plot (\protect\ref{eq:xy2}) for distribution
  (\protect\ref{eq:family}).  Points: Census data for families,
  1947--1994 \cite{history}.  Inset points: Gini coefficient data for
  families from Census \cite{history}.  Inset line: The calculated
  value 3/8 of the Gini coefficient for distribution
  (\protect\ref{eq:family}).}
\label{fig:Lorenz_2}
\end{figure}

Figs.\ \ref{fig:census} and \ref{fig:IRS} demonstrate that the
exponential law (\ref{eq:BG}) fits the individual income distribution
very well.  The Lorenz data for the individual income follow Eq.\
(\ref{eq:Lorenz}) without fitting parameters, and the Gini coefficient
is close to the calculated value 0.5 (Fig.\ \ref{fig:Lorenz_1}).  The
distributions of the individual and family income differ
qualitatively.  The former monotonically increases toward the low end
and has a maximum at zero income (Fig.\ \ref{fig:census}).  The
latter, typically being a sum of two individual incomes, has a maximum
at a finite income and vanishes at zero (Fig.\ \ref{fig:census_2}).
Thus, the inequality of the family income distribution is smaller.
The Lorenz data for families follow the different Eq.\ (\ref{eq:xy2}),
again without fitting parameters, and the Gini coefficient is close to
the smaller calculated value 0.375 (Fig.\ \ref{fig:Lorenz_2}).
Despite different definitions of income by different agencies, the
parameters extracted from the fits (Table \ref{tab:R}) are consistent,
except for line (e).

The qualitative difference between the individual and family income
distributions was emphasized in Ref.\ \cite{SailerWeber}, which split
up joint tax returns of families into individual incomes and combined
separately filed tax returns of married couples into family incomes.
However, Refs.\ \cite{Pub1304} and \cite{Petska} counted only
``individual tax returns'', which also include joint tax returns.
Since only a fraction of families file jointly, we assume that the
latter contribution is small enough not to distort the tax returns
distribution from the individual income distribution significantly.
Similarly, the definition of a family for the data shown in the inset
of Fig.\ \ref{fig:census_2} includes single adults and one-adult
families with children, which constitute 35\% and 10\% of all
families.  The former category is excluded from the definition of a
family for the data \cite{history} shown in Fig.\ \ref{fig:Lorenz_2},
but the latter is included.  Because the latter contribution is
relatively small, we expect the family data in Fig.\
\ref{fig:Lorenz_2} to approximately represent the two-earners
distribution (\ref{eq:family}).  Technically, even for the families
with two (or more) adults shown in Fig.\ \ref{fig:census_2}, we do not
know the exact number of earners.

With all these complications, one should not expect perfect accuracy
for our fits.  There are deviations around zero income in Figs.\
\ref{fig:census}, \ref{fig:IRS}, and \ref{fig:census_2}.  The fits
could be improved there by multiplying the exponential function by a
polynomial.  However, the data may not be accurate at the low end
because of underreporting.  For example, filing a tax return is not
required for incomes below a certain threshold, which ranged in 1999
from \$2,750 to \$14,400 \cite{Pub1040}.  As the Lorenz curves in
Figs.\ \ref{fig:Lorenz_1} and \ref{fig:Lorenz_2} show, there are also
deviations at the high end, possibly where Pareto's power law is
supposed to work.  Nevertheless, the exponential law gives an overall
good description of income distribution for the great majority of the
population.

%%%%%%%%%%%%%%%%%%%%%%%%%%%%%%%%%%%%%%%%%%%%%%%%%%%%%%%%%%%%%%%%%%%%%%
\vspace{-\baselineskip}
\section{Possible origins of exponential distribution}
\label{sec:origin}

The exponential Boltzmann-Gibbs distribution naturally applies to the
quantities that obey a conservation law, such as energy or money
\cite{DY}.  However, there is no fundamental reason why the sum of
incomes (unlike the sum of money) must be conserved.  Indeed, income
is a term in the time derivative of one's money balance (the other
term is spending).  Maybe incomes obey an approximate conservation
law, or somehow the distribution of income is simply proportional to
the distribution of money, which is exponential \cite{DY}.

Another explanation involves hierarchy.  Groups of people have
leaders, which have leaders of a higher order, and so on.  The number
of people decreases geometrically (exponentially) with the
hierarchical level.  If individual income increases linearly with the
hierarchical level, then the income distribution is exponential.
However, if income increases multiplicatively, then the distribution
follows a power law \cite{hierarchy}.  For moderate incomes below
\$100,000, the linear increase may be more realistic.  A similar
scenario is the Bernoulli trials \cite{Feller}, where individuals have
a constant probability of increasing their income by a fixed amount.

\vspace{-\baselineskip}
\begin{acknowledgement}
  We are grateful to D.~Jordan, M.~Weber, and T.~Petska for sending us
  the data from Refs.\ \cite{Pub1304}, \cite{SailerWeber}, and
  \cite{Petska}, to T.~Cranshaw for discussion of income distribution
  in Britain, and to M.~Gubrud for proofreading of the manuscript.
\end{acknowledgement}

%%%%%%%%%%%%%%%%%%%%%%%%%%%%%%%%%%%%%%%%%%%%%%%%%%%%%%%%%%%%%%%%%%%%%%
\vspace{-1.5\baselineskip}

\end{document}